# Pressure Effect on Band Inversion in $AE$Cd$_2$As$_2$

Jonathan M. DeStefano[1,2] and Lin-Lin Wang[2*]

[1]Department of Physics, University of Florida, Gainesville, FL 32608, USA

[2]Ames Laboratory, U.S. Department of Energy, Ames, IA 50011, USA

(February 18, 2021)

## Abstract

Recent studies have predicted that magnetic EuCd$_2$As$_2$ can host several different topological states depending on its magnetic order, including a single pair of Weyl points. Here we report on the bulk properties and band inversion induced by pressure in the non-magnetic analogs $AE$Cd$_2$As$_2$ ($AE$ = Ca, Sr, Ba) as studied with density functional theory calculations. Under ambient pressure we find these compounds are narrow band gap semiconductors, in agreement with experiment. The size of the band gap is dictated by both the increasing ionicity across the $AE$ series which tends to increase the band gap, as well as the larger nearest neighbor Cd-As distance from increasing atomic size which can decrease the band gap because the conduction band edge is an anti-bonding state derived mostly from Cd 5$s$ orbitals. The combination of these two competing effects results in a non-monotonic change of the band gap size across the $AE$ series with SrCd$_2$As$_2$ having the smallest band gap among the three compounds. The application of negative pressure reduces this band gap and causes the band inversion between the Cd 5$s$ and As 4$p$ orbitals along the $\Gamma$-A direction to induce a pair of Dirac points. The topological nature of the Dirac points is then confirmed by finding the closed Fermi arcs on the ($10\bar{1}0$) surface.

* llw@ameslab.gov

# I. Introduction

Symmetry-protected topological states[1-4] have greatly expanded our understanding of electronic band structures in condensed matter physics and materials science. There have been a lot of efforts in classification[5-12] of different topological states and search for topological materials[13-15] with different topological features. Among them, the Dirac semimetal (DSM), such as $Na_3Bi$[16] and $Cd_3As_2$[17], stands out as being a highly tunable topological state which hosts a four-fold degenerated Dirac point between two inverted bands which is protected by the combination of time-reversal and crystalline symmetries. The lifting of such a protected band crossing and degeneracy due to lowering symmetry can result in either a topological insulator (TI) or a Weyl semimetal (WSM). The key ingredient is band inversion. Besides searching for new DSMs, inducing topological phase transitions in systems, such as $ZrTe_5$[18, 19], has also been realized by tuning band inversion with pressure[20], temperature[21], strain[22] and even coherent phonons[23-25].

Compounds in the $CaAl_2Si_2$-type structure of space group 164 (*P-3m1*) have been studied for dilute magnetic semiconductors[26], thermoelectrics[27, 28], and possible hosts for superconductivity[29]. More recently $EuCd_2As_2$ has been predicted as a topological magnetic semimetal, being antiferromagnetic (AFM) DSM[30] or ferromagnetic (FM) WSM with a single pair of Weyl points[31, 32] along the *Γ-A* direction, depending on the magnetic configurations[33, 34]. These studies indicate that this crystal structure can host a variety of electronic structures and multiple physically interesting states. In contrast to the semimetallic $EuCd_2As_2$, for the non-magnetic version of *AE*$Cd_2As_2$, where alkali earth (*AE*) is Ca, Sr and Ba, recent experiments found they are narrow gap semiconductors. Measurements of resistivity vs temperature shows the transport band gap is 0.2 and 0.4 eV for $SrCd_2As_2$[26] and $BaCd_2As_2$[27], respectively. However, the calculated band gap in density functional theory (DFT) varies. Regular generalized gradient approximation (GGA) exchange-correlation functionals, such as PBE[35], without spin-orbit coupling (SOC) has been used[26] to give a band gap agreeing with experiment. But for meta-GGA, such as the modified Becke-Johnson[36, 37] (mBJ) exchange-correlation functionals with SOC, the band gap is overestimated to be as large as 1.0 eV[27, 28].

Here using PBEsol[38] and SOC included to fully relax the lattice parameters of *AE*$Cd_2As_2$, we find a narrow band gap of 0.175, 0.095 and 0.199 eV for $CaCd_2As_2$, $SrCd_2As_2$ and $BaCd_2As_2$, respectively. Although they are underestimated by about half as compared to experimental values, the trend of SrCd$_2$As$_2$ having a band gap about half that of BaCd$_2$As$_2$ agrees with the available experimental data[26, 27]. We also predict that CaCd$_2$As$_2$ has a larger band gap than SrCd$_2$As$_2$. We explain the non-monotonic change of band gap size across the *AE* series by identifying the valence and conduction band edge as being mostly derived from As 4*p* orbitals for bonding states and Cd 5*s* orbitals for anti-bonding states, respectively. The size of the band gap is affected by the competing factors of both the ionicity of *AE* and the Cd-As nearest neighbor distance. Across the *AE* series from Ca to Sr and then to Ba, the increasing ionicity will increase the band gap. But because the conduction band edge is derived from Cd 5*s* orbitals as a part of the anti-bonding states, the increasing Cd-As nearest neighbor distance across the series with larger lattice constant also tends to decrease the band gap. Thus, combining these two effects, the band gap first decreases and then increases when moving from Ca to Sr and Ba with a minimum at SrCd$_2$As$_2$. With magnetism in EuCd$_2$As$_2$ to further shift bands from exchange splitting, the band gap closes with band inversion and EuCd$_2$As$_2$ becomes a magnetic semimetal with a very small Fermi surface[31]. In this paper, we focus on the bulk properties and pressure effect on the band inversion in the non-magnetic *AE*Cd$_2$As$_2$. Because the band inversion is between Cd 5*s* and As 4*p* orbitals near the *Γ* point which is sensitive to the Cd-As nearest neighbor distance, we find that negative pressure is an effective knob to tune such band inversion along *Γ-A* direction to form a pair of Dirac points. We confirmed the Dirac point by calculating the topological surface states, which give the closed Fermi arcs on the side ($10\bar{1}0$) surface connecting the projections of the pair of Dirac points.

## II. Computational Methods

All the electronic band structures, total energies, forces and stresses of *AE*Cd$_2$As$_2$ have been calculated in DFT[39, 40] using PBEsol[38] and PBE[35] exchange-correlation functional as implemented in VASP[41, 42] with a plane-wave basis set and projector augmented wave (PAW) method[43]. Spin-orbit coupling effect is included in all the calculations. We use a *Γ*-centered Monkhorst-Pack[44] (11×11×7) *k*-point mesh for the trigonal unit cell of five atoms. The kinetic energy cutoff to expand the plane-wave coefficients is 415 eV for PAW with semi-core orbitals of *AE* included as valence. The convergence with respect to *k*-point mesh was carefully checked, with total energy converged below 1 *m*eV/cell. For ionic relaxation, the absolute magnitude of force on each atom is reduced below 0.01 eV/Å with the target pressure values. For phonon calculation, the small displacement method in Phonopy[45] has been used together with VASP in a (3×3×2) supercell. A tight-binding model based on the maximally localized Wannier functions (MLWFs)[46, 47] was constructed to reproduce closely the band structure including SOC in the range of ±1eV around the Fermi energy ($E_F$). Then the spectral functions and Fermi surface (FS) of a semi-infinite surface were calculated using the surface Green's function methods[48-51] as implemented in WannierTools[52].

## III. Results and Discussion

### A. Crystal structure and bulk properties

The crystal structure of *AE*Cd$_2$As$_2$ (Fig.1(a) inset) has a trigonal unit cell in space group 164 (*P-3m1*). The *AE* atoms at *1a* position have a simple hexagonal lattice. The Cd and As atoms at *2b* positions form Cd$_2$As$_2$ layers along the *c*-axis and are separated by the *AE* layers. The stacking sequence is -*AE*-As-Cd-Cd-As- along the *c*-axis. The Cd atom is locally coordinated by four As atoms in almost a tetrahedra. The nearest neighbor Cd-As distance is with the three equivalent As atoms on the equilateral triangle base, while the larger Cd-As distance along the *c*-axis is for the next nearest neighbor. The crystal structure has the inversion center at the *AE* site and the inversion symmetry is not broken by the application of external hydrostatic pressure.

To study the bulk properties of *AE*Cd$_2$As$_2$ under hydrostatic pressure, we have calculated total energy ($E_{tot}$) versus volume ($V$) with both PBE+SOC and PBEsol+SOC. For nine volumes near the equilibrium, both the shape of unit cell and the internal coordinates are optimized. The equilibrium volume ($V_0$), bulk modulus ($B_0$) and its derivative with respect to pressure ($B_0'$) were obtained by fitting $E_{tot}$ versus $V$ (see Fig.1 (a)-(c)) to the Burch-Murnaghan[53] equation of state which has the following form:

$$E_{tot}(V) = E_0 + \frac{9V_0 B_0}{16}\left\{\left[\left(\frac{V_0}{V}\right)^{\frac{2}{3}} - 1\right]^3 B_0' + \left[\left(\frac{V_0}{V}\right)^{\frac{2}{3}} - 1\right]^2 \left[6 - 4\left(\frac{V_0}{V}\right)^{\frac{2}{3}}\right]\right\} \quad (1)$$

The results are listed in Table 1. The lattice constants and equilibrium volume increase across the *AE* series from Ca to Sr and then to Ba, as expected for the increasing atomic/ionic size. Correspondingly, the bulk modulus monotonically decreases across the *AE* series because there is more open space in the unit cell with increasing size, allowing easier compression of the electron density for the same number of electrons.

The fully relaxed lattice parameters of *AE*Cd$_2$As$_2$ under ambient pressure using both PBE+SOC and PBEsol+SOC are plotted against the experimental lattice parameters in Fig.1(d). The in-plane *a* lattice parameters calculated with PBEsol+SOC agree very well with the experimental data[54, 55] being within 1%, while those from PBE+SOC are overestimated by more than 1%. The calculated *c* parameters tend to be smaller than experiment with PBEsol+SOC and larger with PBE+SOC. Due to the better agreement with experiment on the in-plane *a* lattice parameter, which is important for the nearest neighbor Cd-As distance, PBEsol+SOC has been used for the rest of the calculations on electronic band structures.

## B. Bulk band structures and band gap

The electronic band structures under ambient pressure of *AE*Cd$_2$As$_2$ calculated with PBEsol+SOC using the fully relaxed lattice parameters are presented in Fig.2(a)-(c). There is a sizable gap between the top valence and bottom conduction band along the high symmetry directions except near the $\Gamma$ point for all the three compounds. The calculated direct band gaps ($\Delta$) across the series are listed in Table 1 and also plotted in Fig.2(d) together with the available experimental data. Usually across the *AE* series from Ca to Sr and Ba, the band gap should monotonically increase due to the increased ionicity in chemical bonds because *AE* is more likely to lose electrons. However, as plotted in Fig.2(d), the change of the band gap across the series is not monotonic with the Sr compound having the smallest band gap of 0.095 eV. This can be explained by noticing the nature of the band gap and the specific orbital content of the bottom of conduction band along the $\Gamma$-*A* direction, which can affect the band gap size when considering together with the change of lattice constants across the *AE* series.

Using SrCd$_2$As$_2$ as an example, the density of states (DOS) projected on Sr 5*s*, Cd 5*s* and As 4*p* orbitals are plotted in Fig.3(a). The valence bands are clearly dominated by As 4*p* orbitals. The Sr 5*s* spreads out across the energy range except for the absence near the top of the valence bands and has a large contribution to the conduction bands, which corresponds to a charge transfer from Sr to the Cd$_2$As$_2$ layers. In contrast, Cd 5*s* has a strong hybridization with As 4*p* orbitals forming bonding and anti-bonding states. The bonding states have a narrow range from −5.0 to −4.0 eV near the bottom of the valence bands and the anti-bonding states are from 0.0 to 4.0 eV in the conduction bands, respectively. The bottom of the conduction bands, i.e., the band edge along the $\Gamma$-*A* direction, is the lowest anti-bonding states just above the $E_F$ and mostly derived from Cd 5*s* orbitals, as indicated by the green arrow in Fig.3(a). The character of the conduction band edge just above the $E_F$ can also be seen after zoomed along the *Γ-A* direction in Fig.3(b) with the orbital content of Cd 5*s* being highlighted in green.

Across the *AE* series from Ca to Sr and Ba, besides the increasing ionicity for more ionic bonding to increase the band gap, the larger lattice constants from the increasing atomic size elongates the nearest neighbor Cd-As distance and reduces the energy separation between bonding and anti-bonding states, which can decrease the band gap. Thus, combining these two factors of the opposite tendency, a minimal band gap size of 0.095 eV is realized at the middle of the *AE* series for SrCd$_2$As$_2$ as shown in Fig.2(d).

In comparison to the experimental transport band gap of 0.2 eV for SrCd$_2$As$_2$[26] and 0.4 eV for BaCd$_2$As$_2$[27], our PBEsol+SOC calculated band gap of 0.095 and 0.199 eV are about twice as small. The underestimation of band gaps is a well-known problem for regular GGA exchange-correlation functionals. However, our results agree with the trend in experimental data that BaCd$_2$As$_2$ has about twice as large a band gap than SrCd$_2$As$_2$. From our calculations, we predict that CaCd$_2$As$_2$ also has a larger band gap than SrCd$_2$As$_2$. In contrast, meta-GGA exchange-correlation functionals, such as the mBJ[36, 37], which is good for band gaps of many materials, gives unusually large band gaps for these *AE*Cd$_2$As$_2$ compounds, all about 1.0 eV[27, 28]. The same behavior is also found in our calculations. The hybrid exchange-correlation functionals, such as HSE06[56], which include part of the exact exchange, also gives a much larger band gap than experiment, for example 0.7 eV for SrCd$_2$As$_2$ with experimental lattice parameters from our calculations. The reason for overestimating the band gap of these *AE*Cd$_2$As$_2$ compounds with meta-GGA and hybrid exchange-correlation functionals requires further studies.

**C. Band inversion under negative pressure and topological features**

After establishing that the conduction band edge along the *Γ-A* direction are anti-bonding states derived from Cd 5*s* orbitals as shown in Fig.3(b) and discussed above, next we will discuss the band inversion and pressure effect. The band inversion in *AE*Cd$_2$As$_2$ is along the *Γ-A* direction between Cd 5*s* and As 4*p* orbitals, which is sensitive to the nearest neighbor Cd-As distance and thus pressure. To assess the effect of pressure on band inversion, we choose SrCd$_2$As$_2$ because it has the smallest band gap among the three compounds, and fully relax the lattice under different hydrostatic pressures in the PBEsol+SOC calculations. Because the conduction band derived from Cd 5s just above the $E_F$ is part of the anti-bonding states, a positive pressure which reduces the nearest neighbor Cd-As distance will push the bottom conduction band toward higher energy resulting in a larger band gap. Thus, a negative pressure is needed to reduce the band gap and induce the band inversion. As shown in Fig.3(c), with $P=-1.6$ GPa, the band gap is reduced to a few $m$eV. With a more negative pressure at $P=-1.8$ GPa, the band gap becomes inverted with the crossing point along $\Gamma$-$A$ direction, of which the degeneracy is protected by the 3-fold rotation symmetry and thus a Dirac point. The energy-momentum of this pair of Dirac points are ($E_F$+6 $m$eV; 0.000, 0.000, ±0.012 Å$^{-1}$).

To confirm the topological properties of the Dirac points, we have constructed the maximally localized Wannier functions from the PBEsol+SOC band structure and calculated the surface states for SrCd$_2$As$_2$ under the pressure of $P=-1.8$ GPa. As shown in Fig.4(a), when projected on the basal plane (0001), the projections of the pair of Dirac points overlap with each other and there are no topological surface states. In contrast, when projected to the side plane (10$\bar{1}$0) in Fig.4(b), the projections of the Dirac points are separated along the $\bar{Z}$-$\bar{\Gamma}$-$\bar{Z}$ direction with connecting topological surface states. With a surface Fermi surface at the energy of $E_F$+6 $m$eV, the topological surface states form closed Fermi arcs on the (10$\bar{1}$0) as shown in Fig.4(c). The converging points of the Fermi arcs are discontinuous points from the projection of Dirac points, agreeing with other DSMs such as Na$_3$Bi[16] and Cd$_3$As$_2$[17], as well as the AFM-DSM EuCd$_2$As$_2$[30]. The green arrows show the in-plane spin polarization arising from the spin-momentum locking.

One common question about a structure under hydrostatic pressure is whether it remains dynamically stable or not. To ensure that our predicted Dirac semimetal under negative pressure is dynamically stable, we have calculated the phonon band dispersion and density of states for SrCd$_2$As$_2$ under −1.8 GPa. As plotted in Fig.5(a), all the phonon eigenmodes are positive spanning from 0.0 to 5.5 THz. In Fig.5(b), the phonon DOS are also projected on atomic species. The lower range from 0.0 to 2.6 THz is contributed mostly from the modes associated with Cd. The middle range from 2.6 to 4.0 THz is dominated by Sr including the mode with flat dispersion at 2.7 THz. In contrast, the higher range from 4.0 to 5.5 THz is mostly from the modes associated with As. The phonon calculation clearly shows that our predicted Dirac semimetal of SrCd$_2$As$_2$ under −1.8 GPa is dynamically stable without negative phonon eigenmodes.

For the other two compounds of $CaCd_2As_2$ and $BaCd_2As_2$, we have also calculated the critical pressure ($P_c$) to induce the band inversion and Dirac points as listed in Table 1. Because of the non-monotonic trend of the band gap size across the series, $CaCd_2As_2$ and $BaCd_2As_2$ require more negative $P_c$ at −2.9 and −2.7 GPa, respectively, to realize the band inversion. Unlike positive hydrostatic pressures that can be applied in a diamond anvil cell, effective negative pressures can be realized by alloying with elements from the same group of larger atomic size. We propose that alloying Sb at the As site is the most promising, because of both the larger atomic size and less electronegativity of Sb than As, which are beneficial to induce band inversion. Recently, Weyl semimetal of Sb compounds such as $EuCd_2Sb_2$ have been reported[57].

Lastly, to demonstrate that the band structure topology of the Dirac point in $SrCd_2As_2$ under negative pressure is robustly protected by symmetry, we have increased the negative pressure in our calculation, which is more conveniently done in theory than experiment. The band inversion is near the $\Gamma$ point with the Dirac point along the $\Gamma$-$A$ direction being protected by the 3-fold rotational symmetry. When the pressure is increased to −3.6 GPa in Fig.6(a) and then to −7.2 GPa in Fig.6(b), there are energy shifts in the valence bands at $\Gamma$ and $A$ points, and the band dispersion slopes along the $\Gamma$-$A$ direction are decreased because of the increased lattice constants under higher negative pressure. However, the inverted band order at $\Gamma$ point between the top valence and bottom conduction bands does not change. The band inversion actually increases as shown by the shift of the Dirac point toward the $A$ point. The Dirac point is still protected by the 3-fold rotation, which is intact under hydrostatic pressure. The other reason for the persistence of Dirac point over a large range of negative pressure is from the large band gap at the $A$ point, which is reduced from 1.0 eV (Fig.2(b)) to 0.8 eV then 0.6 eV when the pressure is increased from 0 GPa to −3.6 GPa and then to −7.2 GPa. The combination of these factors results in the robust Dirac point in $SrCd_2As_2$ under a large range of negative hydrostatic pressure.

## IV. Conclusion

In conclusion, we have studied the bulk properties, electronic band structure and band inversion under negative pressure of $AE$Cd$_2$As$_2$ ($AE$=Ca, Sr, Ba) by first-principles calculations based on density functional theory (DFT). Using PBEsol exchange-correlation functional with spin-orbit coupling included, we find the good agreement to experimental data on the relaxed lattice parameters. Although the bulk moduli and lattice constants have a monotonic change across the *AE* series, we find that the size of the narrow band gap does not have a monotonic increase. Besides the increasing ionicity to increase the band gap, the increasing lattice constant coupled with the conduction band edge along the $\varGamma$-$A$ direction being anti-bonding state of Cd 5s orbitals can decrease the band gap. The combination of these two competing effects across the *AE* series results in a non-monotonic change of the band gap size with $SrCd_2As_2$ having the smallest band gap. The trend of the band gap change agrees with available data on $SrCd_2As_2$ and $BaCd_2As_2$, and we predict that $CaCd_2As_2$ has a larger band gap than $SrCd_2As_2$. We find that the band inversion in $AECd_2As_2$ is along the $\varGamma$-$A$ direction between Cd 5$s$ and As 4$p$ orbitals, which is sensitive to the nearest neighbor Cd-As distance and thus pressure. Because the conduction band edge along the $\varGamma$-$A$ direction is of anti-bonding states, negative pressure can reduce the band gap and induce the band inversion to form Dirac points. We confirmed the Dirac point by calculating the topological surface states, which give the closed Fermi arcs on the side ($10\bar{1}0$) surface connecting the projections of the pair of Dirac points. We propose that alloying can be used to realize the effective negative pressure to induce the band inversion and Dirac points in these compounds.

## Acknowledgements

We acknowledge the helpful discussion with and the support from Paul C. Canfield. JMD was supported by the U.S. Department of Energy Office of Science, Science Undergraduate Laboratory Internships (SULI) program under its contract with Iowa State University, Contract No. DE-AC02-07CH11358. This work was supported as part of the Center for the Advancement of Topological Semimetals, an Energy Frontier Research Center funded by the U.S. Department of Energy Office of Science, Office of Basic Energy Sciences through the Ames Laboratory under its Contract No. DE-AC02-07CH11358.

| | Expt. | PBE+SOC | PBEsol+SOC | | | | | Expt. |
|---|---|---|---|---|---|---|---|---|
| | $a$ (Å) $c$ (Å) $z_{As}$ $z_{Cd}$ | $a$ (Å) $c$ (Å) $z_{As}$ $z_{Cd}$ | $a$ (Å) $c$ (Å) $z_{As}$ $z_{Cd}$ | $B_0$ (GPa) | $B'_0$ | $P_c$ (GPa) | $\Delta$ (eV) | $\Delta$ (eV) |
| CaCd$_2$As$_2$ | 4.391[a] 7.184 0.2382 0.6331 | 4.441 7.217 0.2355 0.6340 | 4.369 7.073 0.2343 0.6334 | 55.4 | 4.54 | −2.9 | 0.175 | -- |
| SrCd$_2$As$_2$ | 4.460[b] 7.420 0.278 0.63 | 4.512 7.477 0.2497 0.6325 | 4.435 7.344 0.2491 0.6322 | 53.2 | 4.56 | −1.8 | 0.095 | 0.2[c] |
| BaCd$_2$As$_2$ | 4.513[b] 7.674 0.278 0.63 | 4.577 7.745 0.2623 0.6300 | 4.500 7.619 0.2612 0.6298 | 51.4 | 4.64 | −2.7 | 0.199 | 0.4[d] |

a Ref.[54], b Ref.[55], c Ref[26]. and d Ref[27].

Table 1. Fully-relaxed lattice parameters ($a$, $c$, $z_{As}$ and $z_{Cd}$) with PBEsol+SOC and PBE+SOC in comparison to experiments. The parameters fit to Burch-Murnaghan equation of state with PBEsol+SOC for bulk modulus ($B_0$) and its derivative ($B'_0$). The calculated band gap ($\Delta$) with PBEsol+SOC in fully-relaxed lattices in comparison to experiments, as well as the critical pressure ($P_c$) to induce band inversion. (See main text for the band gaps calculated with other exchange-correlation functionals).

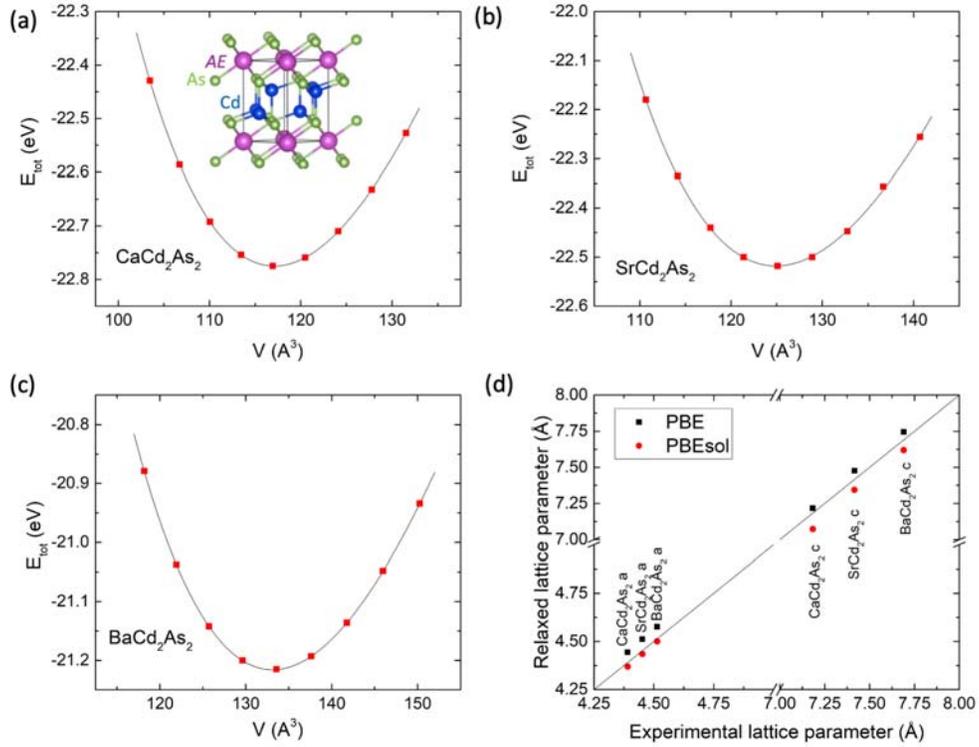

Figure 1. Total energy ($E_{tot}$) as a function of volume ($V$) calculated with PBEsol+SOC fit to the Burch-Murnaghan equation of state for (a) CaCd$_2$As$_2$ (inset: crystal structure of $AE$Cd$_2$As$_2$ in space group 164), (b) SrCd$_2$As$_2$, and (c) BaCd$_2$As$_2$. (d) calculated lattice parameters ($a$ and $c$) using PBE+SOC and PBEsol+SOC compared with experimental lattice parameters for $AE$Cd$_2$As$_2$.

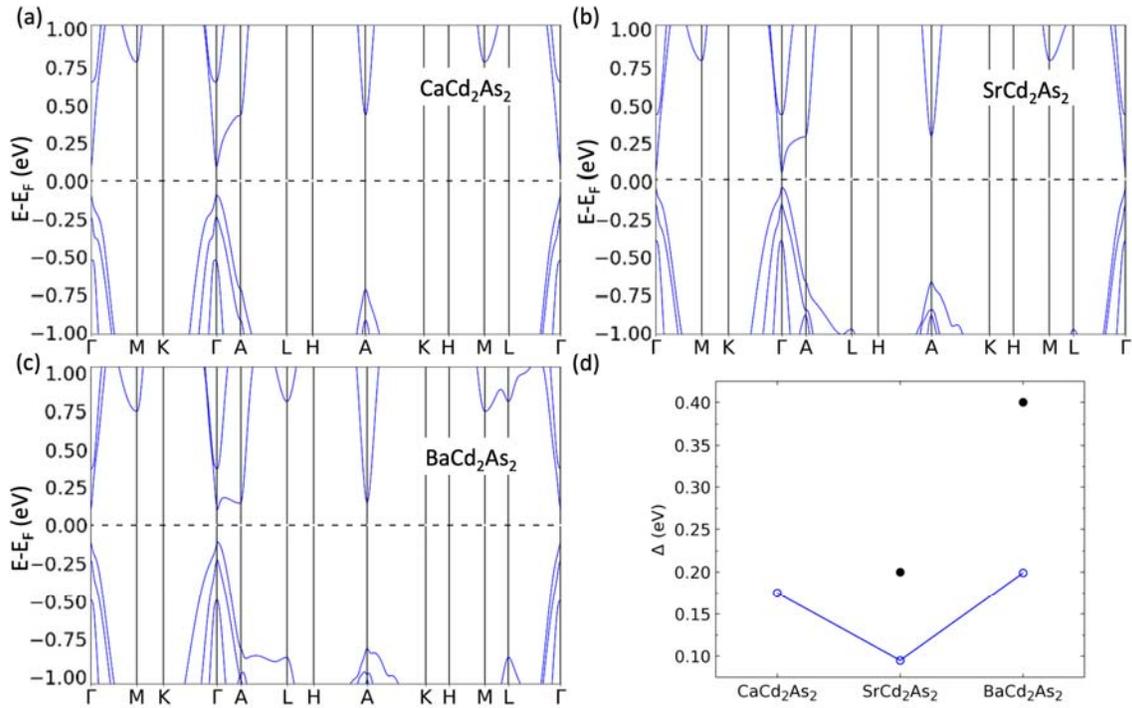

Figure 2. Calculated band structures with PBEsol+SOC using the fully relaxed lattice parameters for (a) CaCd$_2$As$_2$, (b) SrCd$_2$As$_2$, and (c) BaCd$_2$As$_2$. (d) Band gap ($\Delta$) for the $AE$Cd$_2$As$_2$ series with available experimental data in black solid circles.

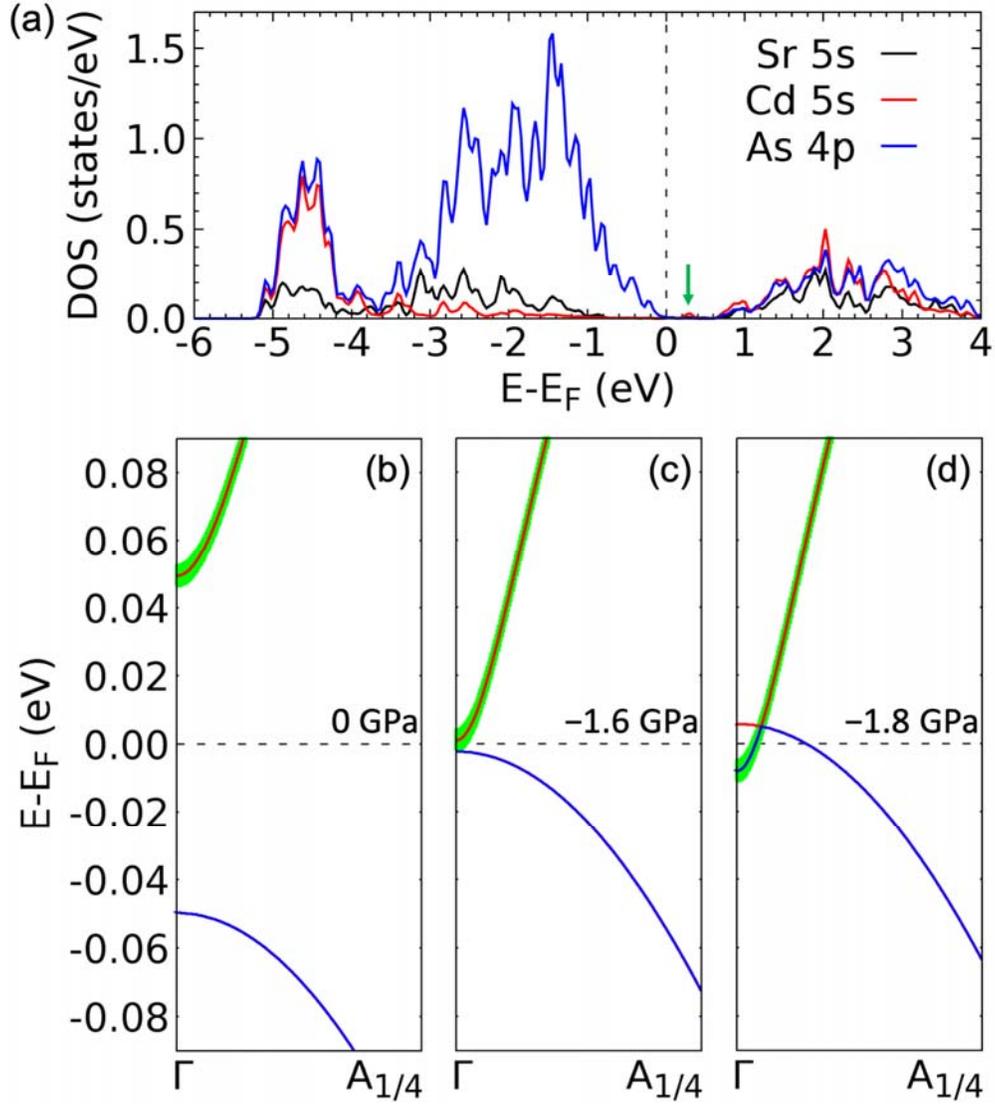

Figure 3. (a) Density of states (DOS) projected on Sr 5*s*, Cd 5*s* and As 4*p* orbitals in SrCd$_2$As$_2$. Green vertical arrow points to the Cd 5*s* band just above the Fermi energy ($E_F$). Panel (b), (c) and (d) are the band structures zoomed along $\Gamma$-*A* direction under the pressure of 0, −1.6 and −1.8 GPa. The top valence (bottom conduction) band is shown in blue (red). The green shadow is the projection on Cd 5*s* orbital.

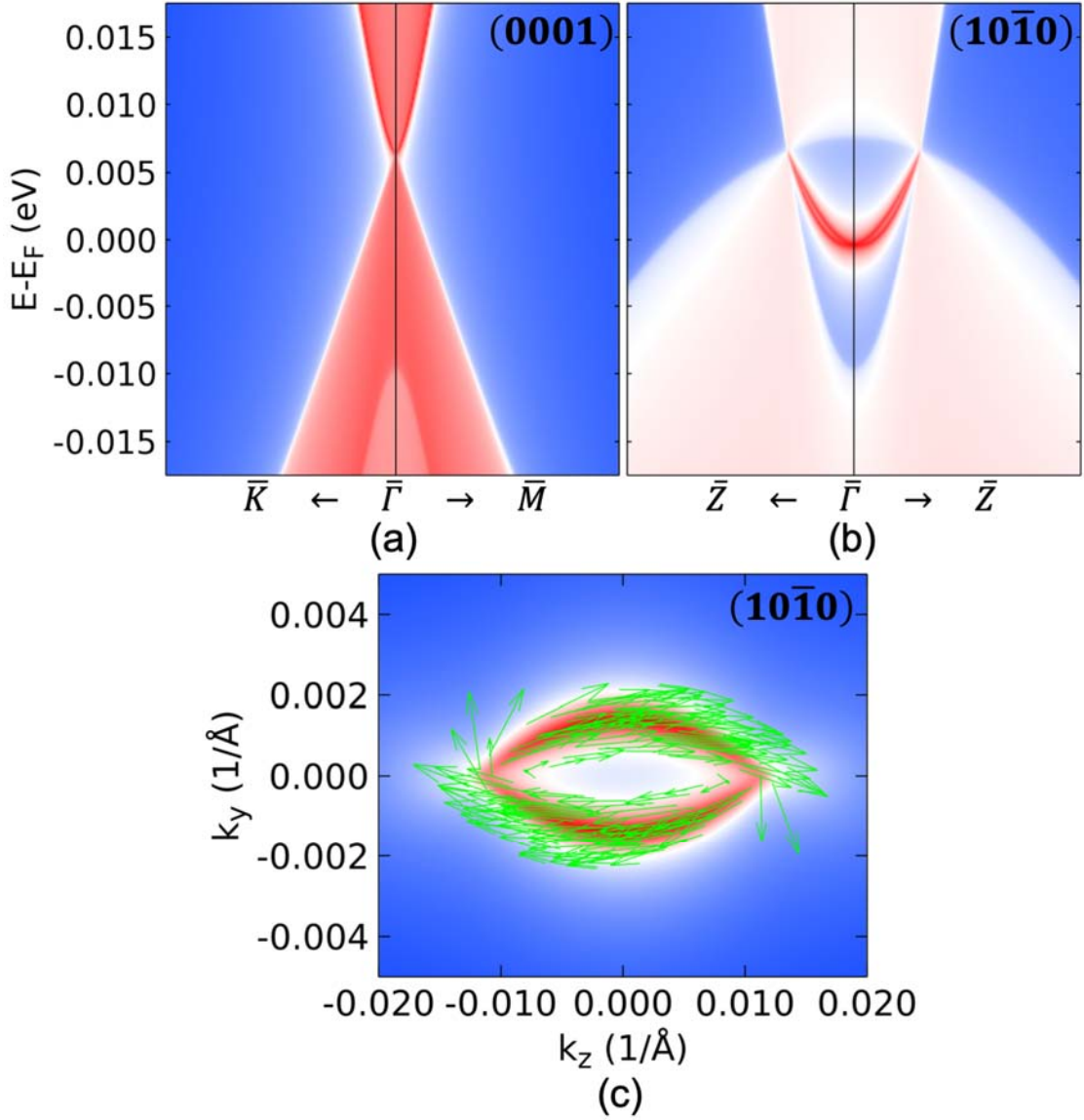

Figure 4. (a) (0001) and (b) (10$\bar{1}$0) surface states of SrCd$_2$As$_2$ under −1.8 GPa with Dirac point from band inversion. (c) 2D Fermi surface (FS) on (10$\bar{1}$0) surface at E$_F$+6 *m*eV. The energy-momentum coordinates of the DP pair are (E$_F$+6 *m*eV; 0.000, 0.000, ±0.012 Å$^{-1}$). Low, medium, and high density of states is indicated by blue, white and red colors, respectively. The green arrows in (c) show the in-plane spin momentum.

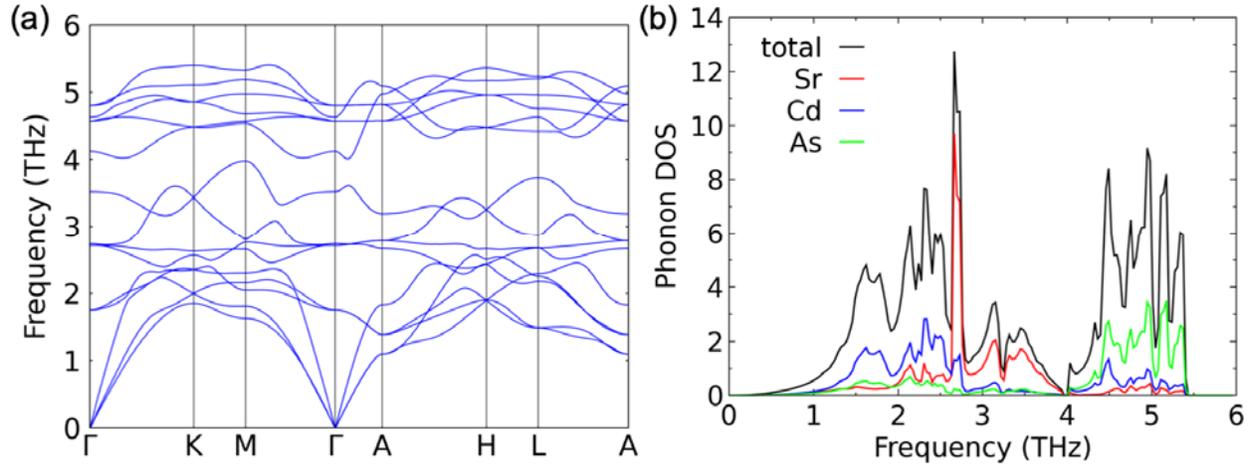

Figure 5. (a) Phonon band dispersion and (b) density of states (DOS) for $SrCd_2As_2$ under $-1.8$ GPa in the Dirac semimetal phase. In (b) phonon DOS are also projected on atomic species.

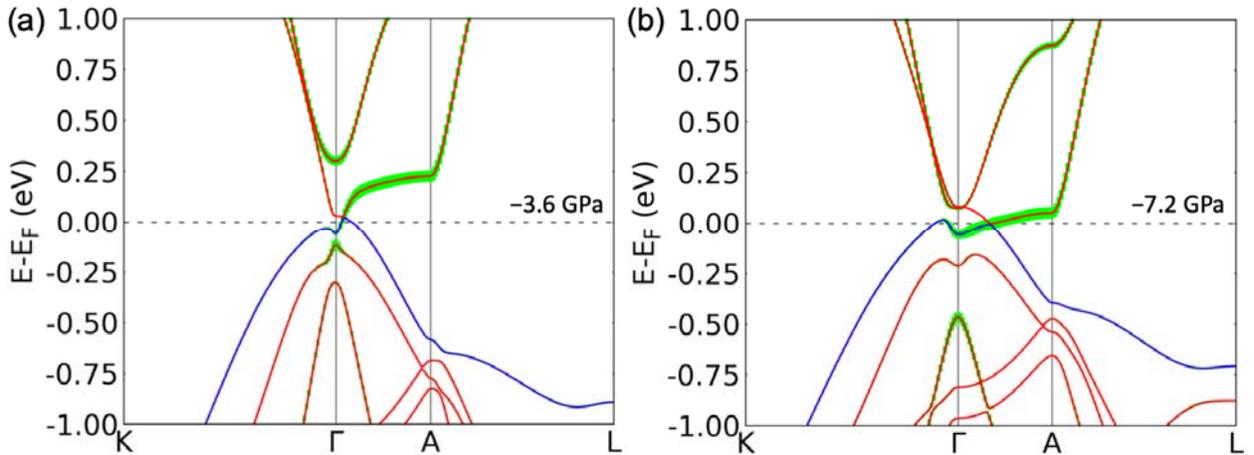

Figure 6. Band structure of $SrCd_2As_2$ under negative pressure of (a) $-3.6$ and (b) $-7.2$ GPa showing the shift of the Dirac point along the $\Gamma$-$A$ direction. The top valence band according to simple band filling is in blue. The green shadow is the projection on Cd $5s$ orbital.